\documentstyle[11pt,newpasp,twoside,epsf]{article}
\def\edcomment#1{\iffalse\marginpar{\raggedright\sl#1\/}\else\relax\fi}
\reversemarginpar

\begin{document}
\title{HETE-2 Localization and Observations of the Short, Hard Gamma-Ray 
Burst GRB020531}

\author{D. Q. Lamb$^1$, G. R. Ricker$^2$, J.-L. Atteia$^3$, K.
Hurley$^4$, N. Kawai$^5$, Y. Shirasaki$^6$, T. Sakamoto$^5$, T.
Tamagawa$^7$, T. Donaghy$^1$, C.~Graziani$^1$, C. Barraud$^3$,
J.-F.~Olive$^3$, A. Yoshida$^8$, K. Torii$^7$, E.~E.~Fenimore$^9$,
M.~Galassi$^9$, R.~Vanderspek$^2$ and the HETE-2 Science Team}

\affil{$^1$U. Chicago, $^2$MIT, $^3$CESR, $^4$UC Berkeley, $^5$Titech, 
$^6$NAOJ, $^7$RIKEN, $^8$Aoyama, $^9$LANL}

\begin{abstract}

The HETE-2 FREGATE and WXM instruments detected a short, hard GRB at
00:26:18.72 UT on 31 May 2002.  A preliminary localization was reported
as a GCN Position Notice 88 min after the burst, and a refined
localization was disseminated 123 minutes later.  An IPN localization
of the burst was reported 18 hours after the GRB, and a refined IPN
localization was disseminated $\approx$5 days after the burst.  The
final IPN localization, disseminated on 25 July 2002, is a 
diamond-shaped region centered on RA=15$^{\rm h}$ 15$^{\rm m}$
11.18$^{\rm s}$, Dec=-19$^\circ$ 24' 27.08" (J2000), and has an area of
$\approx$9 square arcminutes (99.7\% confidence region).  The prompt
localization of the burst by HETE-2, coupled with the refinement of the
localization by the IPN, made possible the most sensitive follow-up
observations to date of a short, hard GRB at radio, optical, and X-ray
wavelengths.  The time history of GRB020531 at high ($>30$ keV)
energies consists of a short, intense spike followed by a much less
intense secondary peak, which is characteristic of many short, hard
bursts.  The duration of the burst increases with decreasing energy and
the spectrum of the burst evolves from hard to soft, behaviors which
are similar to those of long GRBs.  This suggests that short, hard GRBs
are closely related to long GRBs.

\end{abstract}

\section{Introduction}

Gamma-ray bursts appear to fall into two classes: short ($\approx$ 0.2
sec), harder bursts, which comprise 20-25\% of all bursts; and long
($\approx$ 20 sec), softer bursts, which comprise 75-80\% of the total
(Kouveliotou et al. 1993, Lamb, Graziani \& Smith 1993). The spectra of
the two classes of bursts differ: the spectra of the bursts become
softer as the bursts become longer. To date little is known about the
distance to or the nature of the short GRBs, despite extensive
efforts. 

\section{Results}

{\it HETE-2\ } (Atteia et al. 2001, Kawai et al. 2001, Ricker et al.
2001) has detected and localized a short, hard GRB.  Figure 1 (left
panel) shows the final rectangular HETE-2 WXM error box and the final
diamond-shaped IPN error box for this burst, GRB 020531. This
demonstrates that the detection and localization of short, hard GRBs in
the hard x-ray energy band is possible, and has important implications
for Swift and for other future GRB missions.

\begin{figure}
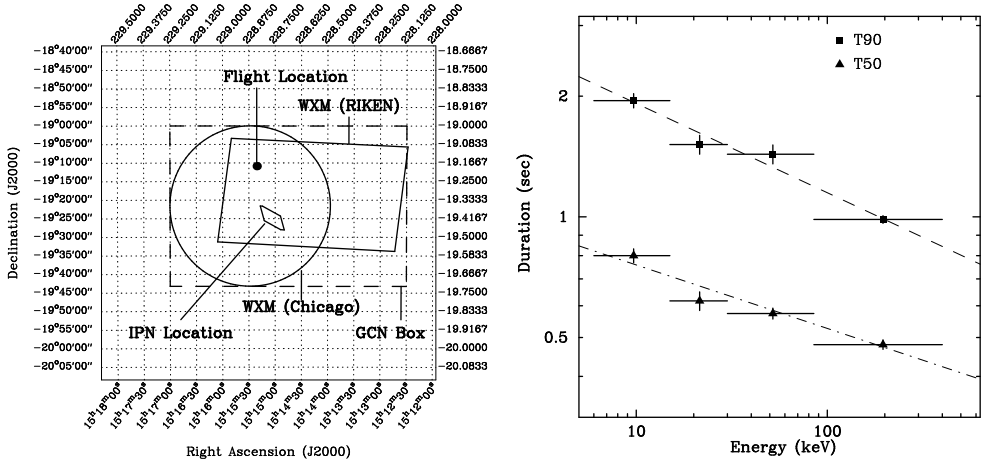

\plotfiddle{lambd3_1.ps}{5.5cm}{0.0}{35.0}{35.0}{-200}{-50}
\plotfiddle{lambd3_2.ps}{0.0cm}{0.0}{35.0}{35.0}{-10}{-75}
\caption{{\it Left} - The final rectangular HETE-2 WXM error box for
GRB020531.  The rectangle completely encloses the Chicago 90\%
confidence region error circle and RIKEN 90\% confidence region error
rectangle (which utilized the same data).  Note that the WXM flight
location lies well inside the confidence region.  Also shown is the
diamond-shaped refined IPN error box for GRB020531, determined by
triangulation using the {\it HETE-2} FREGATE, {\it Ulysses} GRB, and 
{\it Mars Odyssey} HEND data for the burst.  {\it Right} - Duration of
GRB020531 versus energy.  The energy bins have been chosen so that each
contains approximately the same number of photons.  The durations
$t_{50}$ and $t_{90}$ increase with decreasing energy as $E^{-0.16}$
and $E^{-0.22}$, respectively.}
\end{figure}

Figure 2 shows the time history of GRB 020531 in the $85-300$ keV
energy band. The time history of GRB020531 at high energies ($> 30$
keV) consists  of a short, intense spike followed by a much less
intense secondary peak.  Its time history is thus similar to that seen
in many short, hard bursts.  The time history of GRB020531 at low
energies ($< 25$ keV) consists  of two peaks of similar duration
($\approx$ 0.80 sec) and intensity, the first of which corresponds in
time to the short, intense spike seen at high energies. Figure 1 (right
panel) shows that the durations $t_{50}$ and $t_{90}$ increase with
decreasing energy as $E^{-0.16}$ and $E^{-0.22}$, respectively.

The spectrum of the short, intense spike is well described by a single
power-law spectrum with index $\alpha = - 1.16 \pm 0.05$ from 6--400
keV.  Such a steep spectrum is quite unusual, but not unprecedented for
short, hard GRBs (Paciesas et al. 2001).  The steepness of the spectrum
of GRB020531 may explain in part the fact that the WXM on {\it HETE-2}\
was able to detect and localize this particular short, hard burst,
whereas the Wide-Field Cameras on {\it BeppoSAX}\ were ultimately
unsuccessful in detecting or localizing any short, hard GRBs, despite
great efforts.

The spectrum of the secondary peak, which is comparable in duration and
in intensity in the 2--25 keV energy band to the primary peak, is also
well described by a single power-law spectrum, but the spectrum is
softer than the spectrum of the primary peak at the 90\% confidence
level.

The prompt localization of the burst by {\it HETE-2}\ and the anti-Sun
pointing of the {\it HETE-2\ } instruments, coupled with the later
precise localization of the burst by the IPN, allowed rapid follow-up
of the GRB.  Figure 3 shows that these observations have placed much
more severe upper limits on any optical afterglow of a short, hard GRB
than ever before.  These constraints do not rule out the existence of
optical afterglows of short, hard GRBs that are similar to the optical
afterglows of long GRBs.  However, Chandra follow-up observations at
$5$ days found $L_{X}({\rm short})/L_{X}({\rm long}) < 0.01-0.03$
(Butler et al. 2002).

\section{Conclusions} 

The properties of GRB020531 as measured by {\it HETE-2\ } have
different  characteristics at different energies: more complex time
structure at lower energies, increasing duration with decreasing
energy, a power-law spectrum over the 2-400 keV energy range but
spectral softening with  time).  These properties of GRB020531 as
measured by {\it HETE-2} are similar to those of long bursts, and when
taken together with the previously known properties of short, hard GRBs
(similar brightness distributions, $V/V_{\rm max}$ values, angular
distributions) suggest that short, hard GRBs are closely related to
long GRBs.  Chandra follow-up observations at $5$ days found
$L_{X}({\rm short})/L_{X}({\rm long}) < 0.01-0.03$ (Butler et al.
2002).

\begin{figure}
\plotfiddle{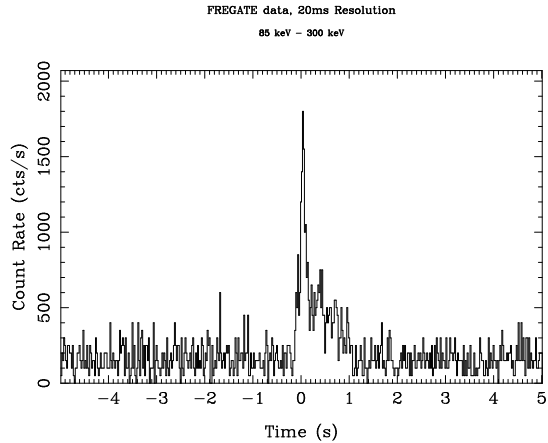}{5.0cm}{270.0}{30.0}{30.0}{-110}{170}
\caption{FREGATE time history of GRB020531 in the $85-300$~keV band, 
binned in 80 msec bins.}
\end{figure}

\begin{figure}
\plotfiddle{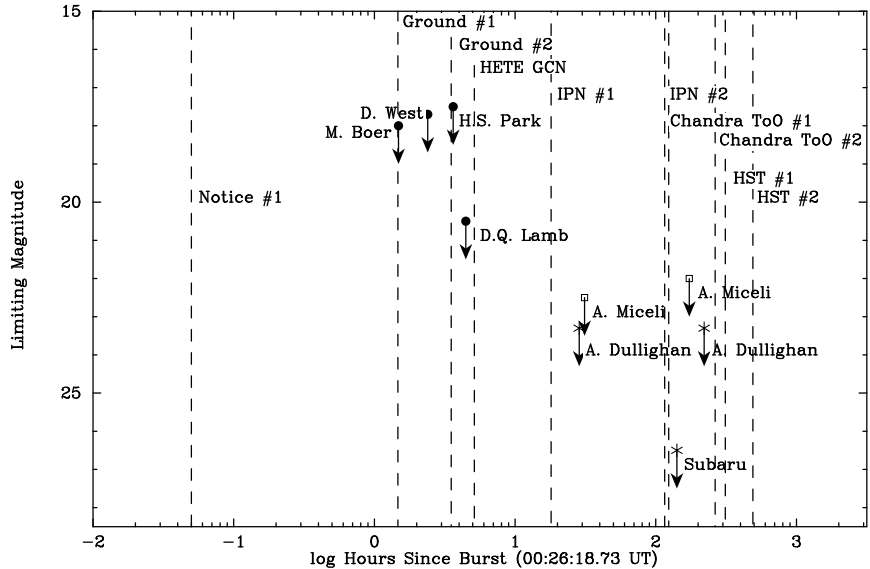}{6.5cm}{270.0}{45.0}{45.0}{-180}{240}
\caption{Limiting magnitudes versus time for any optical afterglow from 
follow-up observations of GRB020531.  The limiting magnitudes shown
include both limits derived using small aperture, large FOV robotic
telescopes (Park et al. 2002, Boer et al. 2002), a small  aperture,
modest FOV telescope (West et al. 2002), and large aperture, small  FOV
telescopes (Lamb et al. 2002, Miceli et al. 2002, Dullighan et al.
2002).}
\end{figure}

\end{document}